\documentclass[twocolumn]{aastex62}
\usepackage{amssymb}
\usepackage{natbib}
\usepackage{hyperref}
\usepackage{color}
\usepackage{pdfcomment}

\newcommand\changed[1]{{#1}}

\received{2018 April 27}
\revised{2018 June 1}
\accepted{2018 June 4}
\submitjournal{PASP}

\shorttitle{Search for Dust Emission from (24) Themis Using the Gemini-North Observatory}
\shortauthors{Hsieh et al.}

\begin{document}

\title{Search for Dust Emission from (24) Themis Using the Gemini-North Telescope\footnote{This is an accepted manuscript version of an article accepted for publication in Publications of the Astronomical Society of the Pacific.  Nether the Astronomical Society of the Pacific nor IOP Publishing Ltd is responsible for any errors or omissions in this version of the manuscript or any version derived from it.}}

\correspondingauthor{Henry Hsieh}
\email{hhsieh@psi.edu}

\author[0000-0001-7225-9271]{Henry H.\ Hsieh}
\affil{Planetary Science Institute, 1700 East Fort Lowell Rd., Suite 106, Tucson, AZ 85719, USA}
\affil{Institute of Astronomy and Astrophysics, Academia Sinica, P.O.\ Box 23-141, Taipei 10617, Taiwan}

\author[0000-0002-4676-2196]{Yoonyoung Kim}
\affiliation{Department of Physics and Astronomy, Seoul National University, Gwanak, Seoul 151-742, Republic of Korea}

\author[0000-0003-0250-9911]{Alan Fitzsimmons}
\affiliation{Astrophysics Research Centre, Queens University Belfast, Belfast BT7 1NN, United Kingdom}

\author{Mark V.\ Sykes}
\affil{Planetary Science Institute, 1700 East Fort Lowell Rd., Suite 106, Tucson, AZ 85719, USA}

\begin{abstract}
We report the results of a search for a dust trail aligned with the orbit plane of the large main-belt asteroid (24) Themis, which has been reported to have water ice frost on its surface.  Observations were obtained with the GMOS instrument on the Gemini-North Observatory in imaging mode, where we used a chip gap to block much of the light from the asteroid, allowing us to take long exposures while avoiding saturation by the object.  No dust trail is detected within 2$'$ of Themis to a 3-$\sigma$ limiting surface brightness magnitude of $\Sigma_{\rm lim}\sim29.7$~mag~arcsec$^{-2}$, as measured along the expected direction of the dust trail. Detailed consideration of dust ejection physics indicates that particles large enough to form a detectable dust trail were unlikely to be ejected as a result of sublimation from an object as large as Themis. We nonetheless demonstrate that our observations would have been capable of detecting faint dust emission as close as 20$''$ from the object, even in a crowded star field.  This approach could be used to conduct future searches for sublimation-generated dust emission from Themis or other large asteroids closer to perihelion than was done in this work.  It would also be useful for deep imaging of collisionally generated dust emission from large asteroids at times when the visibility of dust features are expected to be maximized, such as during orbit plane crossings, during close approaches to the Earth, or following detected impact events.
\end{abstract}



\section{Introduction}\label{section:intro}

\subsection{Background}\label{subsection:background}

Dust emission from solar system bodies is most commonly associated with comets from the outer solar system, but there are situations in which asteroids in the inner solar system may be expected and, in some cases, have even been observed to also eject dust.  Collisional erosion of asteroid surfaces is one means by which asteroids can be sources of dust and debris in their orbital vicinity. Because impact velocities in the main asteroid belt are $\sim$5~km~s$^{-1}$ \citep{bottke1994_astcollisionvelocities}, significant energy is transferred to ejecta during an impact event, a significant fraction of which will escape the gravitational field of an asteroid.  Dust is also produced in the catastrophic disruptions of parent asteroids believed to create asteroid families \citep{nesvorny2003_dustbands,nesvorny2006_astdustbands,nesvorny2008_beagle}, taking the form of dust bands that have been identified in data obtained by the Infrared Astronomical Satellite \citep{low1984_infraredcirrus,dermott1984_irasdustbands,sykes1990_zodiacaldustbands}. 

Asteroids on cometary orbits (ACOs), which appear inactive but have comet-like orbits \citep[as defined by having Tisserand parameter values of $T_J<3$;][]{kresak1972_tisserand}, are often believed to be dormant comets \citep[e.g.,][]{fernandez2001_neoalbedos,fernandez2005_neoalbedos,licandro2006_acos,licandro2008_acos1}, and so may have the potential to exhibit new dust emission due to reactivation or residual dust emission from earlier, more active periods \citep[e.g.,][]{mommert2014_donquixote}.  Active asteroids have asteroid-like orbits (as defined by having $T_J>3$) but exhibit comet-like dust emission due to various mechanisms \citep{jewitt2015_actvasts_ast4}, and consist of disrupted asteroids, which exhibit activity due to processes such as impact disruptions or rotational destabilization \citep[cf.][]{hsieh2012_scheila}, and main-belt comets (MBCs), which exhibit activity determined to be due to the sublimation of volatile ice \citep{hsieh2006_mbcs,snodgrass2017_mbcs}.

While most currently known active asteroids are relatively small \citep[effective nucleus radii of $r_n\lesssim5$~km; references in][]{jewitt2015_actvasts_ast4}, two large disrupted asteroids have been observed, namely (596) Scheila and (493) Griseldis, which have $r_n=79.9$~km and $r_n=20.8$~km, respectively \citep{mainzer2016_neowise}.  
The dust emission observed for each object is believed to have been caused by a non-catastrophic impact on each body \citep{jewitt2011_scheila,bodewits2011_scheila,ishiguro2011_scheila2,tholen2015_griseldis}.  Scheila and Griseldis were both quite bright when they were discovered to be active, with apparent $V$-band magnitudes of $m_V=14.3$~mag and $m_V=15.7$~mag (according to the JPL Horizons online ephemeris generator\footnote{{\tt https://ssd.jpl.nasa.gov/horizons.cgi}}), respectively.  Main-belt dwarf planet (1) Ceres \citep[$r_n=467.6$~km;][]{carry2008_ceres}, is sometimes considered an active asteroid due to the fact that it has been observed to exhibit water vapor outgassing \citep{kuppers2014_ceres}, but visible dust emission has never been observed from this body.  No large ($r_n>5$~km) MBCs or MBC candidates seen to exhibit observable dust emission are currently known.

\subsection{Asteroid Ice}\label{subsection:themis}

Volatile material is known to have existed and to still exist in asteroids.  Meteorites linked to the asteroid belt contain aqueously altered minerals, indicating that liquid water was once present in main belt objects \citep[e.g.,][]{hiroi1996_cgbfasteroids,burbine1998_gclassasts,keil2000_asteroidthermalalteration}. Even long before the discovery of water vapor outgassing discussed above (Section~\ref{subsection:background}), Ceres was thought to possess present-day surface water ice \citep{lebofsky1981_ceresstructuralwater,vernazza2005_ceresvesta}, which has now been directly detected by Dawn \citep{combe2016_cereswater}.  
Spectroscopic evidence of water ice frost has also been reported for main-belt asteroid (24) Themis \citep{rivkin2010_themis,campins2010_themis}, fellow Themis family asteroid (90) Antiope \citep{hargrove2015_antiope}, and other asteroids \citep{takir2012_3micron}.
Volatile material in main-belt asteroids is a subject of great interest in astrobiology, given dynamical studies indicating that objects from the region of the solar system occupied by the present-day main asteroid belt, or at least similar to objects currently occupying the present-day main belt, could have played a significant role in the primordial delivery of water to the terrestrial planets \citep[][and references within]{morbidelli2000_earthwater,raymond2004_earthwater,raymond2017_waterorigin,obrien2006_earthwater,obrien2018_waterdelivery}.

The Themis asteroid family
has come to be of particular interest in studies of asteroid ice in recent years.  Along with the spectroscopic detections of surface ice on Themis and Antiope discussed above, at least three MBCs are likely members of the family \citep{hsieh2004_133p,hsieh2011_176p,hsieh2012_288p}.  Since asteroid family members are believed to be compositionally similar, this evidence of probable ice on multiple Themis family members suggests that ice could be widespread in the family.

While the presence of ice on MBCs is inferred indirectly from their dust emission activity, no spectroscopic detections of water ice on a MBC exhibiting visible dust emission have ever been made.  The nuclei of the known MBCs are simply too small ($r_n\ll5$~km) and faint ($m_V\gg20$~mag when inactive) to obtain spectroscopic observations sensitive enough to detect any water ice on their surfaces that may be present.  Conversely, detectable cometary activity has never been observed for Themis and Antiope despite the reported spectroscopic detections of water ice on their surfaces.  Attempts to directly detect sublimation from Themis have been made but have been unsuccessful thus far \citep{jewitt2012_themiscybele,mckay2017_themisceres}. However, given that MBCs exhibit visible dust emission without exhibiting detectable outgassing \citep[see references in][]{snodgrass2017_mbcs}, the same could be true for Themis.
A major challenge to searching for dust emission from Themis, though, is that it is difficult to efficiently obtain useful deep-imaging observations of such a bright object with a large telescope without quickly saturating the detector and accumulating intractable levels of scattered light.

\changed{Photometric analysis (i.e., searching for photometric enhancements that could indicate the presence of unresolved ejected material) is an alternative method of detecting dust emission activity from an object \citep[e.g.,][]{bus1988_chiron,tholen1988_chiron,cikota2014_activeasts,hsieh2015_324p}.  Use of this method, however, requires that the amount of ejected dust contained within the seeing disk of an object has a total scattering surface area comprising a detectable fraction of the scattering surface area of the object itself.
For very large objects like Themis, this requirement may represent an implausibly large amount of material to be ejected. Higher quality observations with greater photometric precision could improve sensitivity to smaller amounts of ejected material, but even then, this method would be limited by the precision of 
the predictions of Themis's magnitude when inactive, as well as the need for photometric excesses to exceed photometric uncertainties due to rotational variations and aspect angle effects in order to be conclusively attributable to ejected material.}

In this paper, we present a search for \changed{visible} dust emission associated with Themis, where our observations and data reduction techniques were specially designed to address the challenges of performing deep imaging of bright asteroids using relatively simple observing techniques (i.e., suitable for queue mode observing) and instrumentation. While we did not detect any dust emission with these particular observations, we suggest that this approach could be used in the future to perform additional searches for faint activity associated with Themis at different points in its orbit, and also faint activity associated with other bright asteroids in general.

\section{Observations}

A primary consideration in the design of our observations to search for dust emission from Themis was the asteroid's extreme brightness, meaning that any long exposure by a large (8-m-class) telescope would quickly saturate.  Bleeding of excess charge, as well as extensive scattered light, would then obscure any faint dust features from view, particularly if the largest column density of dust particles is close to the object.  On the other hand, a long series of extremely short exposures to avoid saturating individual exposures would be inefficient both to obtain and process.  To address these challenges, we devised a method to observe Themis using one of the chip gaps of a multi-element mosaic imager to block the majority of the light from the asteroid, allowing us to take much longer exposures than would be feasible otherwise.

Another consideration in the design of our observations was timing.  While dust emission activity due to sublimation is likely to be strongest near perihelion, or perhaps shortly afterwards \citep[cf.][]{hsieh2011_238p}, dust ejected early in an object's active period is expected to be concentrated near the source object.  This is because shortly after activity begins, ejected particles do not have time to travel very far, unless they are very small (e.g., $\mu$m-scale or smaller), in which case they disperse quickly and rapidly become too diffuse to be visible at large distances from the object.  For an extremely bright asteroid like Themis, though, dust features close to the object are difficult to detect due to interference from scattered light from the asteroid itself.

At longer times after the start of dust emission, large dust particles (e.g., mm-scale or larger, which disperse much more slowly in response to solar radiation pressure and Keplerian shear) can form a dust trail that is relatively long-lasting and extends to large distances from the source object, avoiding the large amount of scattered light expected near the object.
These trails are commonly seen in associated with short-period comets \citep{sykes1992_cometdusttrails,reach2007_cometdebristrails}, and are the type of dust trail that we aimed to search for with the observations of Themis presented here.


\begin{figure*}
\includegraphics[height=3.4in]{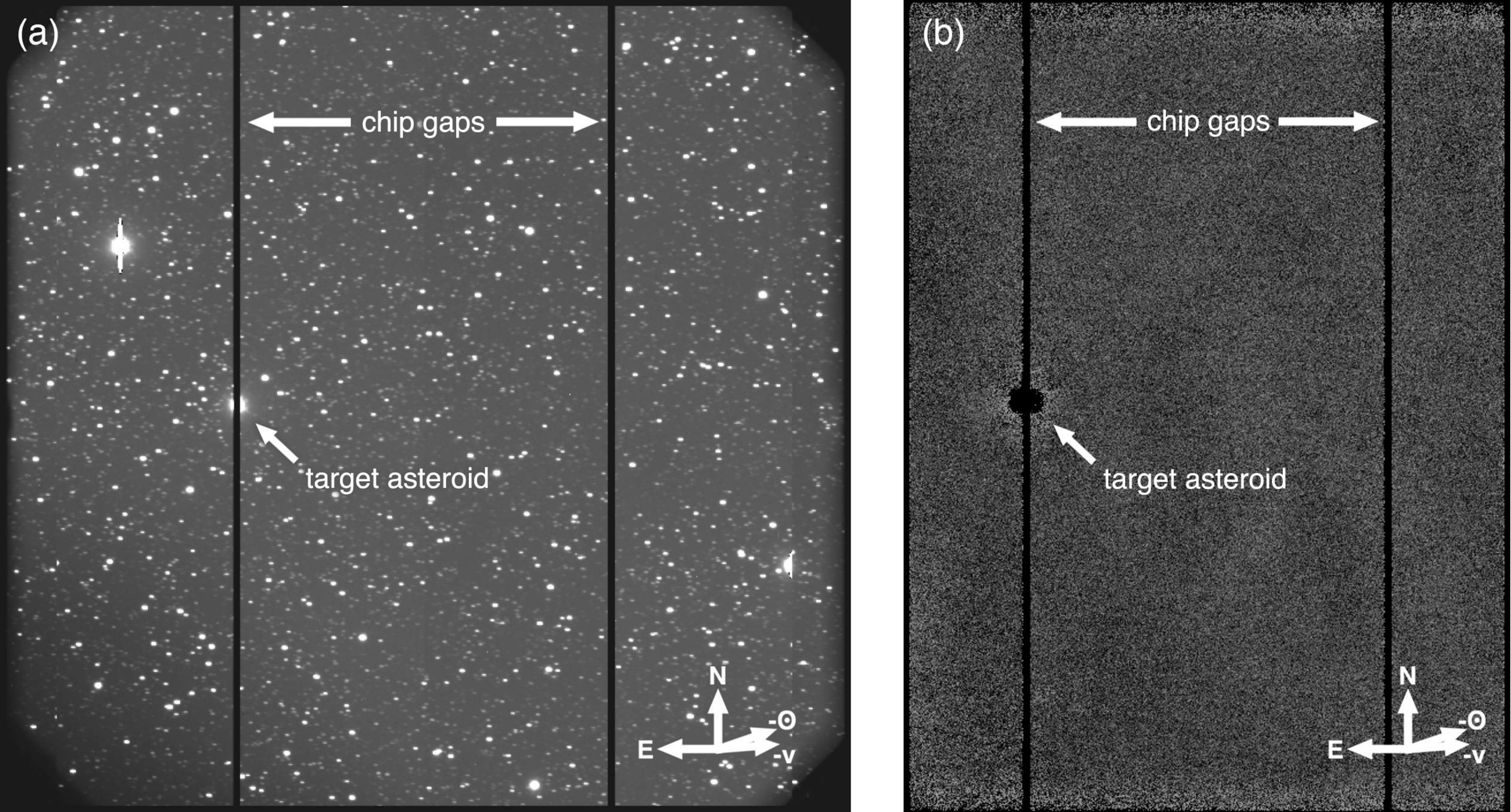}
\caption{(a) A single 180~s exposure of (24) Themis from Gemini-N on 2015 June 9 showing the positions of the chip gaps and the asteroid.
(b) Cleaned composite image (where field stars have been removed) constructed from all data listed in Table~\ref{table:obslog} (2~hrs of total effective exposure time). The locations of the target asteroid and the chip gaps, north (N), east (E), the antisolar direction ($-\odot$), and the negative heliocentric velocity vector ($-v$), as projected on the sky, are marked in both panels.  The central chip area in both panels is $2\farcm4$ wide and $5\farcm0$ high.}
\label{figure:single_composite}
\end{figure*}

We obtained queue observations of Themis on 2015 June 9 and 2015 June 12 (Table~\ref{table:obslog}) using the 8.1~m Gemini North telescope (Program GN-2015A-FT-19) on Mauna Kea in Hawaii.  
At the time of our observations, Themis was at a true anomaly of $\nu$$\,\sim\,$117$^{\circ}$, similar to the orbit positions at which other MBCs have been observed to show long, faint dust trails detectable with observations similar to our planned Gemini observations of Themis \citep[i.e., 238P/Read and 324P/La Sagra;][]{hsieh2009_238p,hsieh2014_324p}.

Observations were made using the imaging mode of the Gemini Multi-Object Spectrograph \citep[GMOS; image scale of $0.1454''$~pixel$^{-1}$;][]{hook2004_gmos} and a Sloan $r'$ filter.  
Using accurate knowledge of the position and non-sidereal rate of motion of Themis from Horizons\footnote{\tt http://ssd.jpl.nasa.gov/horizons.cgi}, we were able to design our observations such that minimal effort would be required on the part of queue observers to maintain the object's position in the GMOS chip gap once the telescope was positioned properly at the start of observations.

Later, this same accurate knowledge of Themis's position and non-sidereal rate of motion allowed us to accurately combine our data even though Themis itself was not directly imaged by our observations.  Using this technique, assuming $1\farcs0$ (FWHM) seeing conditions in $r'$-band, placing Themis in the $2\farcs8$ chip gap between the GMOS e2V DD chips would allow us to avoid $\gtrsim\,$95\% of the flux from the asteroid \citep{diego1985_stellarimageprofiles}, minimizing the potential for saturation, with the remaining flux in the wings of the object's surface brightness profile reaching the detector on either side of the chip gap.  Observations comprising one hour of total effective exposure time were obtained on each night, giving us a total of two hours of total on-source integration time.

\begin{table*}
\caption{Observation Log}
\label{table:obslog}
\begin{tabular}{lcccccDDDDDDD} 
\hline\hline
 \multicolumn{1}{c}{UT Date} &
 {Telescope} &
 {$N$$^a$} &
 {$t$$^b$} &
 {am$_{\rm mid}$$^c$} &
 {$\theta_s$$^d$} &
 \multicolumn2c{$\nu$$^e$} &
 \multicolumn2c{$R$$^f$} &
 \multicolumn2c{$\Delta$$^g$} &
 \multicolumn2c{$\alpha$$^h$} &
 \multicolumn2c{PA$_{-\odot}$$^i$} &
 \multicolumn2c{PA$_{-v}$$^j$} &
 \multicolumn2c{$\Delta t$$^k$} \\
\hline
\decimals
2013 October 31 & {\it Perihelion} & ... & ...  & ...   & ... &   0.0 & 2.740 & 3.146 & 17.8 & 292.6 & 292.7 &   0 \\
2015 June 09    & Gemini           &  20 & 3600 & 1.408 & 0.8 & 117.3 & 3.275 & 2.261 &  0.9 & 289.0 & 274.5 & 586 \\
2015 June 12    & Gemini           &  20 & 3600 & 1.481 & 0.7 & 117.8 & 3.278 & 2.263 &  0.3 &  46.9 & 274.7 & 589 \\
\hline\hline
\multicolumn{13}{l}{$^a$ Number of exposures.} \\
\multicolumn{13}{l}{$^b$ Total exposure time, in seconds.} \\
\multicolumn{13}{l}{$^c$ Airmass at midpoint of observations.} \\
\multicolumn{13}{l}{$^d$ FWHM seeing, in arcsec.} \\
\multicolumn{13}{l}{$^e$ True anomaly, in degrees.} \\
\multicolumn{13}{l}{$^f$ Heliocentric distance, in AU.} \\
\multicolumn{13}{l}{$^g$ Geocentric distance, in AU.} \\
\multicolumn{13}{l}{$^h$ Solar phase angle, in degrees.} \\
\multicolumn{13}{l}{$^i$ Position angle of the antisolar vector, in degrees east of north.} \\
\multicolumn{13}{l}{$^j$ Position angle of the negative heliocentric velocity vector, in degrees east of north.} \\
\multicolumn{13}{l}{$^k$ Days past perihelion.}
\end{tabular}
\end{table*}

\section{Data Reduction}

We performed standard bias subtraction and flat-field reduction for all data using sky flats constructed from dithered science images.
Flux calibration was conducted using field star photometry from the Pan-STARRS1 survey catalog \citep{tonry2012_ps1,schlafly2012_ps1,magnier2013_ps1}.
To enable precise positional offsetting of our images for constructing composite images, we derived independent astrometric solutions for each image using the 2MASS point source catalog \citep{cutri2003_2masssources}, ultimately attaining an astrometric precision of $0.2''$ for each image.

During our observations, Themis was passing through a dense star field near the galactic plane (galactic latitude of $b$$\,\sim\,$8$^{\circ}$), meaning that careful data reduction was needed to search for potential dust emission features against this crowded background.  As such, we first masked data within 15$''$ of Themis's expected position in each snapshot in order to avoid excess flux from the asteroid, and then created a star-aligned composite image excluding masked pixels and CCD gaps.  We then used the Hotpants image subtraction software package \citep{becker2015_hotpants}\footnote{\tt https://github.com/acbecker/hotpants} and WCSTools/remap astrometric alignment software \citep{mink2011_wcstools}\footnote{\tt http://tdc-www.harvard.edu/wcstools/remap.html} to align and resample, convolve, and normalize the resulting background source template to match each individual image for optimal image subtraction, and subtracted each customized template from individual images.  For further image refinement, we applied masks to saturated pixels and misaligned remnants in individual subtracted images, using SExtractor's ``{\tt check image}'' \citep{bertin1996_sextractor} as a mask image (after careful visual inspection) to remove all sources except the sky background and any faint dust structures.  Individual subtracted images from both nights (comprising a total of two hours of on-source time) were then shifted and aligned (based on Themis's predicted non-sidereal motion on the sky), and median-averaged to make a new background-subtracted composite image for our search for faint dust structures.

\begin{figure*}
\centering\includegraphics[width=4in]{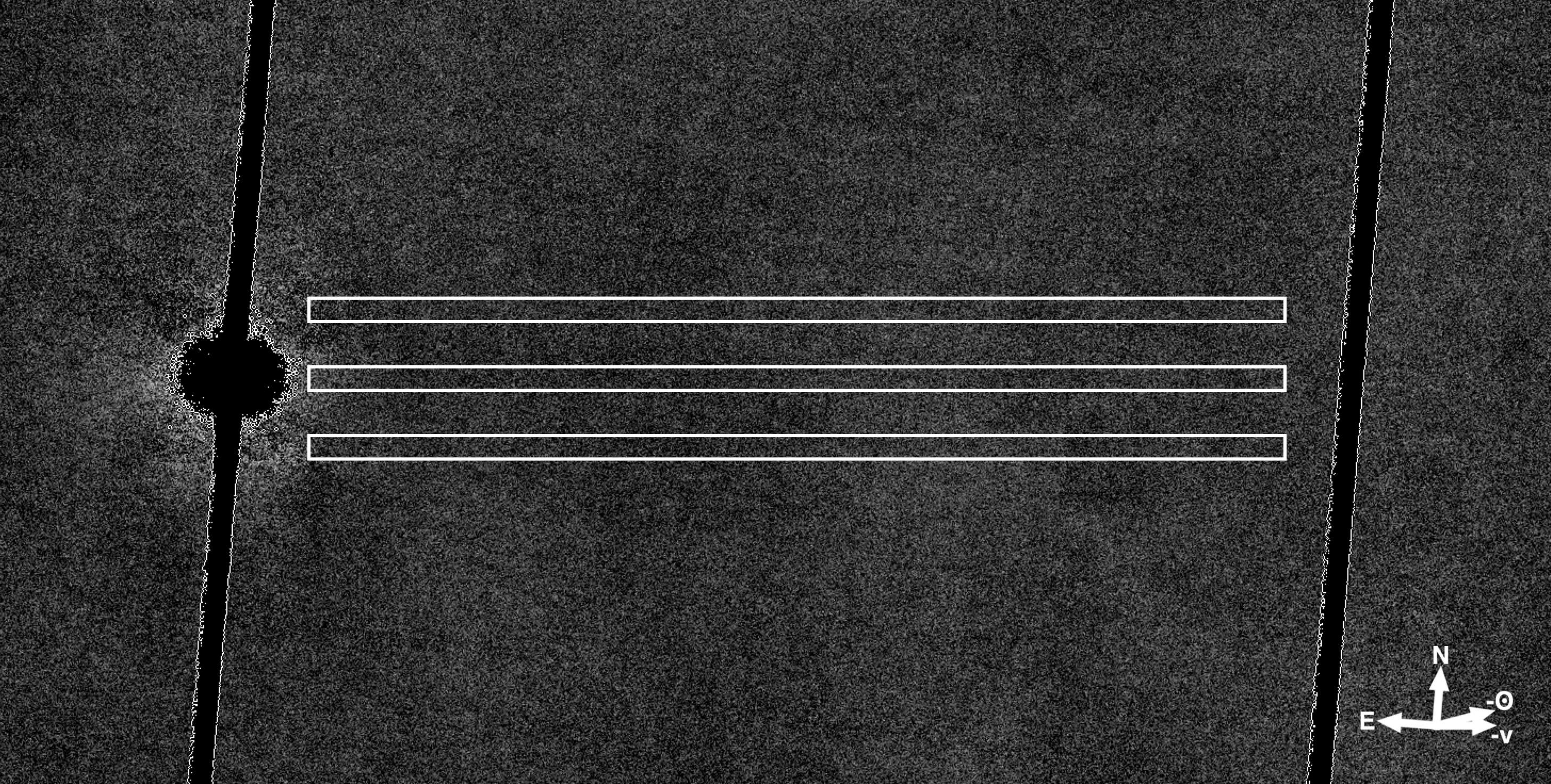}
\caption{Diagram of the positions of the region of sky used to search for dust emission along the expected direction of Themis's dust trail (middle white rectangle) and the regions used for sky background subtraction (top and bottom white rectangles) overlaid on a rotated star-subtracted composite image of our Themis observations.  Themis was located at the center of the large black circle to the left of the image. North (N), east (E), the antisolar direction ($-\odot$), and the negative heliocentric velocity vector ($-v$), as projected on the sky, are marked as labeled.}
\label{figure:composite_image_regions}
\end{figure*}

From a visual inspection of our final composite image including data from both nights (Figure~\ref{figure:single_composite}b), we see no sign of extended dust emission associated with Themis (apart from some residual scattered light from the asteroid itself) down to a 3-$\sigma$ limiting magnitude of $\Sigma_{\rm lim}$$\,\sim\,$28.0~mag~arcsec$^{-2}$, as measured over 1~arcsec$^2$.  To perform a more quantitative search for a dust trail aligned with Themis's orbit plane as projected on the sky, as was seen for 238P and 324P \citep{hsieh2009_238p,hsieh2014_324p} and predicted for Themis using syndyne-synchrone dust modeling analysis \citep[after][]{finson1968_cometdustmodeling1}, we first rotate our final composite image of Themis to make the expected direction of the trail horizontal in the image frame and then measure the surface brightness of the image within a 3$''$-wide strip extending to the west (i.e., in the expected direction of the dust trail) using 5$''$-long rectangular apertures.  We also perform the same measurements for strips of sky 10$''$ North and South of the expected trail in order to measure nearby sky background levels.  A diagram of these regions overlaid on our rotated composite image of Themis is shown in Figure~\ref{figure:composite_image_regions}.

\begin{figure*}
\centering\includegraphics[width=4.5in]{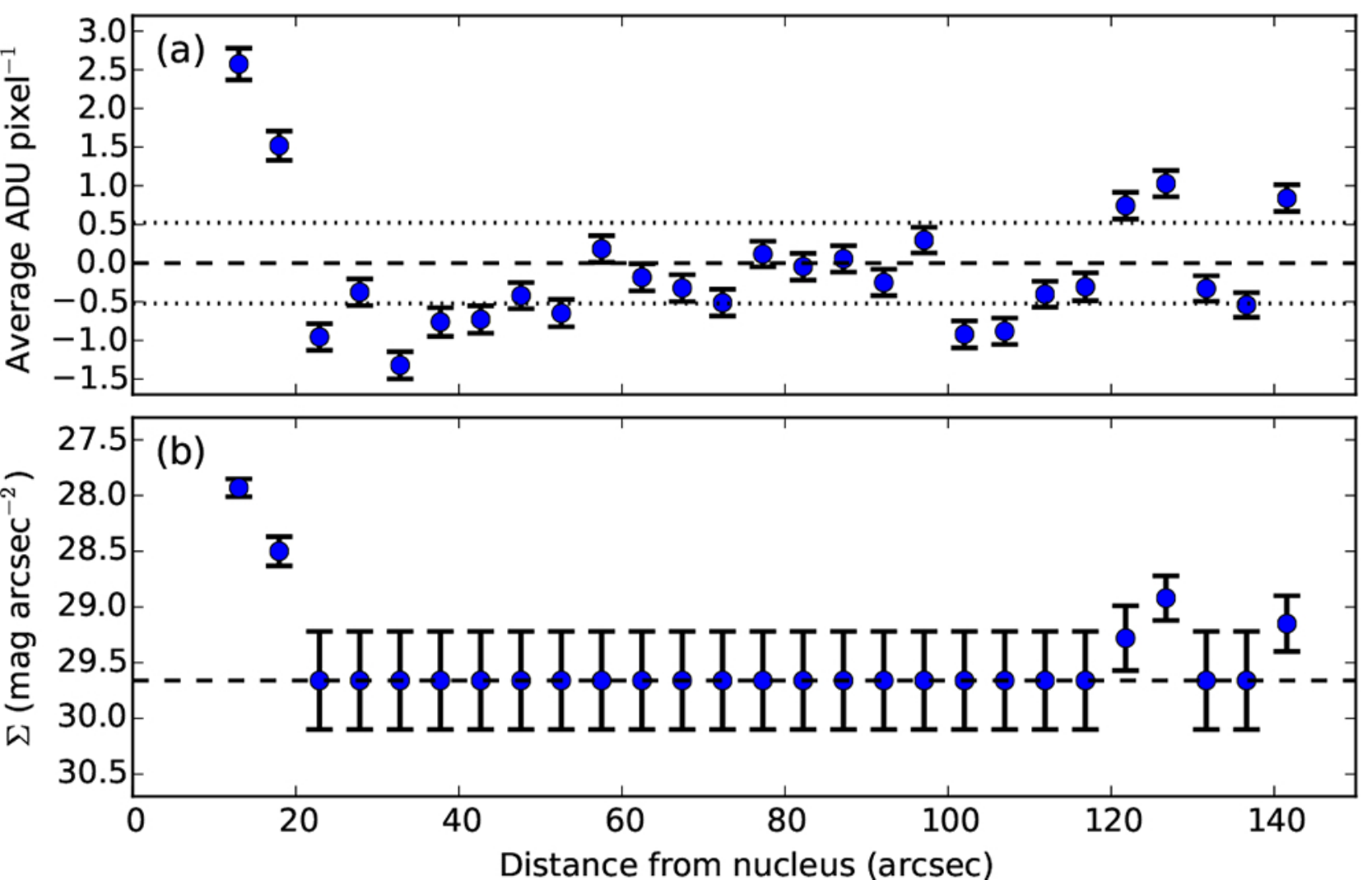}
\caption{(a) Plot of net flux per pixel (in ADU) for 5$''$-long apertures along a 3$''$-wide strip placed along the expected extent of the Themis dust trail in the composite image constructed from data from both nights of observations (schematically shown in Figure~\ref{figure:composite_image_regions}).  Error bars indicating 1-$\sigma$ uncertainties are shown, while points contained within the dotted lines are within 3-$\sigma$ of the sky level. (b) Plot of equivalent surface brightnesses in magnitudes per arcsec$^{2}$ for the aperture fluxes in (a), where magnitudes for average fluxes less than 3-$\sigma$ above the sky are assigned to the 3-$\sigma$ limiting magnitude of 29.7~mag arcsec$^{-2}$ (as measured over a 3$''$$\,\times\,$5$''$ area) for our observations.}
\label{figure:trail_profile}
\end{figure*}

We then calculate the average of the fluxes measured in the two apertures above and below each trail aperture, and subtract the resulting average local sky brightness from the flux measured in each trail aperture to obtain net fluxes along the expected extent of the trail.  We plot these calculated net fluxes (in counts per pixel within each 3$''$$\,\times\,$5$''$ aperture) in Figure~\ref{figure:trail_profile}a, and equivalent surface brightnesses in magnitudes per arcsec in Figure~\ref{figure:trail_profile}b, where net fluxes less than 3-$\sigma$ above the sky are assigned magnitudes equal to the limiting magnitude of our observations (as computed over a 3$''$$\,\times\,$5$''$ area).  From this analysis, we find no clear evidence of excess flux along the direction of the expected trail up to $\sim\,$2$'$ from our target object down to a limiting magnitude of $\Sigma_{\rm lim}$$\,=\,$29.7 mag~arcsec$^{-2}$.  For comparison, the dust trail detected for MBC 324P/La Sagra at a true anomaly of $\nu$$\,=\,$117$^{\circ}$ in 2011 by the Gemini North telescope was seen to extend $>\,$2$'$ from the comet's nucleus and had a measured surface brightness of $\Sigma_R$$\,\sim\,$27.2~mag~arcsec$^{-2}$ \citep{hsieh2014_324p}.

\section{Discussion}

\subsection{Dust Ejection Physics}

Themis has a radius of $r_n$$\,=\,$91.9$\,\pm\,$5.7~km and an estimated mass of $M$$\,=\,$(5.9$\,\pm\,$1.9)$\times$10$^{18}$~kg \citep{carry2012_astdensities}.  Using
\begin{equation}
v_{\rm esc}=\sqrt{2GM\over r_n}
\end{equation}
to compute the escape velocity of the body, we find an escape velocity of $v_{\rm esc}\sim90$~m~s$^{-1}$.  This is extremely high compared to the dust ejection velocities determined for other MBCs \citep[on the order of meters per second; e.g.,][]{hsieh2004_133p,hsieh2009_238p,hsieh2011_176p,moreno2011_324p,moreno2013_p2012t1,licandro2013_288p} for grain sizes ranging from $\mu$m to cm scales.

\begin{figure}
\includegraphics[width=0.9\columnwidth]{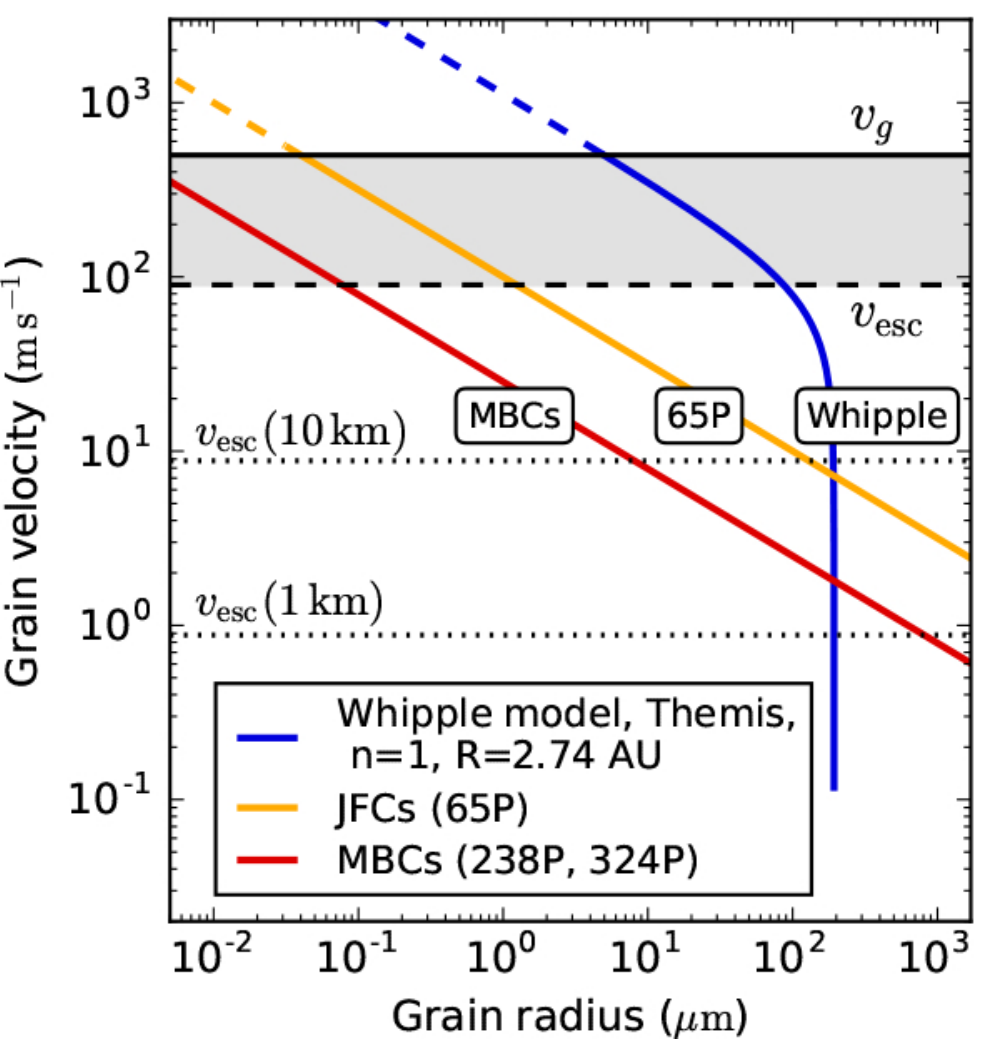}
\caption{Expected dust grain ejection velocities from Themis as a function of particle size,
assuming the Whipple model for dust emission for a ``new'' classical comet at the heliocentric distance of Themis's perihelion (blue line),
the size-velocity relationship for ejected dust grains found for JFC 65P \citep[orange line;][]{ishiguro2007_spcs},
and the size-velocity relationship for ejected dust grains found for previously analyzed MBCs 238P and 324P \citep[red line;][]{hsieh2009_238p,moreno2011_324p}.
A horizontal black solid line marks the assumed mean gas velocity ($v_g=500$~m~s$^{-1}$), while
a horizontal black dashed line marks the escape velocity for Themis ($v_{esc}=90$~m~s$^{-1}$) 
and horizontal black dotted lines mark escape velocities for objects with Themis-like densities ($\rho=1400$~kg~m$^{-3}$) and radii of $r=1$~km ($v_{esc}=0.88$~m~s$^{-1}$) and $r=10$~km ($v_{esc}=8.8$~m~s$^{-1}$), as labeled.
The shaded region indicates the range of velocities over which dust particles exceed Themis's escape velocity, and are therefore ejected and become potentially observable.  Velocities above this range are unphysical because $v_g$ sets an effective maximum velocity for dust particles ejected by gas drag, while particles with velocities below this range cannot escape Themis's gravity and thus never become observable.
}
\label{figure:grainsize_velocity}
\end{figure}

The relationship between particle radius, $a$, and ejection velocity, $v_{ej}$, typically assumed in numerical comet dust ejection models is given by
\begin{equation}
v_{ej}(\beta) \approx v_0 \beta^{1/2}
\label{equation:ejection_velocity}
\end{equation}
where $\beta$ is the dimensionless ratio of the acceleration due to solar radiation pressure to the local acceleration due to gravity for a particle of a given radius (where $\beta\approx1/a$ for $a$ in $\mu$m) and $v_0$ is the reference ejection velocity for a particle with $a=1$~$\mu$m  \citep[e.g.,][]{lisse1998_cobecomets,reach2000_enckedust,ishiguro2007_spcs}.  Assuming $v_0$$\,\sim\,$25~m~s$^{-1}$ as determined for both MBCs 238P/Read and 324P/La Sagra \citep{hsieh2009_238p,moreno2011_324p}, which gives some of the highest ejection velocities determined for a MBC, we find that only dust particles with $\beta$$\,>\,$10 (cf.\ Figure~\ref{figure:grainsize_velocity}), corresponding to $a$$\,<\,$0.1~$\mu$m, would be expected to reach ejection velocities sufficient to escape Themis's gravity.



Sub-$\mu$m particles are not expected to produce the type of long-lasting dust tail that we were searching for with these observations.  Given the large size of Themis, however, higher rates of sublimation than any of the MBCs known to date could be possible in principle, given the possibility of larger subsurface volatile reservoirs.  For example, as mentioned above, (1) Ceres (with a diameter of $\sim$950~km) was observed by {\it Herschel} to exhibit a water vapor production rate of $Q_{H_2O}>10^{26}$~mol~s$^{-1}$, higher than any of the upper limits measured or inferred for known MBCs \citep[e.g.,][]{jewitt2009_259p,jewitt2015_313p1,licandro2011_133p176p,licandro2013_288p,hsieh2012_288p,hsieh2012_324p,hsieh2013_p2012t1,devalborro2012_176pherschel,orourke2013_p2012t1}, perhaps due to either sublimation of near-surface ice (which is likewise a plausible explanation for observed MBC activity) or cryovolcanism powered by heating from long-lived radioisotopes in the deep interior of Ceres \citep{kuppers2014_ceres}.  
In the case of Themis itself, water ice is unstable against sublimation at the calculated equilibrium surface temperatures of the object, yet water frost has been detected across its surface \citep{campins2010_themis}.  This suggests that the detected frost layer was recently deposited at the time of its observations, and perhaps periodically or continuously replenished by mechanisms like ongoing outgassing from the asteroid's interior and re-condensation via cold-trapping \citep{rivkin2010_themis}, in turn, implying the presence of significant subsurface water ice reservoirs.

To get a sense of the upper range of the strength of comet-like sublimation-driven dust emission that we might expect from Themis, we consider the Whipple model \citep{whipple1950_cometmodel1,whipple1951_cometmodel2} for dust emission in classical comets (i.e., from the outer solar system).  In this model, which assumes that sublimation occurs over the entirety of an object's sunlit hemisphere, terminal dust grain velocities are given by
\begin{equation}
v_{\infty} = \left({13{v_g}F\over 12nR[{\rm AU}]^2 a \rho_d H} r_n - {8\pi G\over3}\rho_n r_n^2 \right)^{1/2}
\end{equation}
where $F=1361$~W~m$^{-2}$ is the solar constant, $1/n$ is the efficiency of solar radiation at sublimating cometary ices, $G=6.67\times10^{-11}$~m$^3$~kg$^{-1}$~s$^{-2}$ is the gravitational constant, and $r_n$ is the radius of central body.
Assuming a mean gas velocity of $v_g=500$~m~s$^{-1}$, a bulk asteroid density of $\rho_n=1400$~kg~m$^{-3}$ \citep[characteristic of C-type asteroids like Themis;][]{britt2002_astdensities_ast3}, dust grain densities of $\rho_d=2500$~kg~m$^{-3}$ \citep[characteristic of CI and CM carbonaceous chondrite meteorites which are associated with C-type asteroids;][]{britt2002_astdensities_ast3}, and using Themis's heliocentric distance at perihelion of $R_q=2.74$~AU, and a latent heat of sublimation of $H=2.83\times10^6$~J~kg$^{-1}$, as appropriate for water ice, we obtain
\begin{equation}
v_{\infty} = \left(1.27{1\over n a } - 6.6\times10^3\right)^{1/2}
\end{equation}
We plot this function along with the size-velocity relationship determined for 238P and 324P (Equation~\ref{equation:ejection_velocity}) in Figure~\ref{figure:grainsize_velocity}.  The Whipple model consistently produces larger dust grain velocities than computed for previously observed MBCs, but also includes a steep drop-off in terminal velocity as a critical dust radius is reached.

Setting $v_{\infty}=0$~m~s$^{-1}$, we can express the critical dust particle size above which dust particles are too large to be ejected from a Themis-like asteroid by gas drag as
\begin{equation}
a_c = { 133.2  \over R[{\rm AU}]^2 r_n} {1\over n } \approx {200\over n} ~\mu{\rm m}
\end{equation}
For a ``new'' comet (i.e., $n=1$), only particles with $a<200$~$\mu$m ($\beta>5\times10^{-3}$) can be ejected.  This result reflects the maximum activity strength that we would expect as a result of sublimation across the entire surface of Themis, where $a_c$ will decrease for lower sublimation efficiencies (i.e., larger $n$).
For comparison, using values appropriate for (62) Erato ($r_n=53.5$~km; $R_q=2.61$~AU), another large Themis family asteroid, we find $a_c\approx350~n^{-1}$~$\mu$m under the Whipple model.


Applying simple Finson-Probstein modeling \citep{finson1968_cometdustmodeling1}, we find that 200~$\mu$m particles ejected at the time of Themis's perihelion passage would be $\sim\,$1 degree from the nucleus 590 days later (at the time of our observations), and outside our field of view.  To still be visible within 2 arcmin of the asteroid at the time of our observations, 200~$\mu$m particles would have needed to have been ejected fewer than 200 days prior to our observations, or around November 2014, when Themis was at $R=3.03$~AU and $\nu\sim80^{\circ}$, or later.  Compared to the theoretical maximum ejectable particle size for Themis at perihelion ($\nu\sim0^{\circ}$), $a_c$ at $\nu\sim80^{\circ}$ under the full-strength Whipple model is slightly reduced to $a_c\sim$160~$\mu$m (an essentially negligible difference given the expected real-world precision of the critical particle size calculations presented here).  However, the sublimation rate of water is roughly an order of magnitude lower at this distance \citep[cf.][]{hsieh2015_ps1mbcs}, reducing the likelihood of observable levels of dust emission.

To summarize, we have considered the case here of the largest ejectable particles from Themis at perihelion assuming that these particles would be the most likely to contribute to the production of a long-lived dust trail.  Smaller particles could also produce visible dust emission, but only if they were ejected relatively recently with respect to the time of observations.  In the case of the observations presented here, the late timing of our observations relative to Themis's perihelion passage means that any still-visible particles would have had to have been ejected at larger heliocentric distances, at which the sublimation rate of water may have been too low to have produced detectable dust emission.  If Themis is actively emitting dust at any point in its orbit, future observations performed earlier in its orbit (i.e., closer to perihelion) would be more likely to reveal visible dust features.  Similar searches for dust emission associated with other large, potentially icy asteroids should also be performed relatively close to perihelion to maximize the likelihood of detecting visible dust.

\subsection{Potential Alternative Applications}

Besides performing additional searches for dust emission from Themis and other large, potentially icy asteroids near perihelion, the approach described here could also be useful for searches for dust emitted from large asteroids as a result of processes other than sublimation-driven dust emission.  As mentioned above (Section~\ref{subsection:background}), collisional erosion of asteroids could be a significant source of interplanetary dust.
An online impact simulator created by K.\ A.\ Holsapple\footnote{\tt http://keith.aa.washington.edu/craterdata/scaling/ index.htm} demonstrates that for asteroid sizes between 10~km and 1000~km, the total mass of ejecta material created by impactors greater than 1~m in diameter that escape the target asteroid, scaled by the target surface area, monotonically increases with the size of the target asteroid.  This result implies that searches for faint dust associated with the largest asteroids in the asteroid belt could potentially be highly fruitful.

Dust with velocities significantly exceeding the escape velocity of Themis will necessarily form a trail of greater physical extent than the narrower trails associated with sublimation-driven emission of large particles by short-period comets. Such a mechanism may be required to explain the existence of Type II dust trails detected by the Infrared Astronomical Satellite \citep{sykes1988_zodiacalstructures}, which are more than an order of magnitude wider because of the corresponding greater dispersion
in ejection velocities. Searches for such a trail associated with Themis may therefore benefit from the use of facilities with even larger fields of view than GMOS, such as SuprimeCam or HyperSuprimeCam on the Subaru telescope \citep{miyazaki2002_suprimecam}.


In particular, the approach used in this work could be applied to large asteroids in young asteroid families such as the Veritas and Beagle families \citep[e.g.,][]{milani1994_veritas,nesvorny2006_youngfamilies,nesvorny2008_beagle,nesvorny2015_astfam_ast4,pravec2018_astclusters},
to see if 
dust from intra-family collisional erosion (which may occur at a higher rate for young family members, given the more highly similar orbits of young family members compared to those of other asteroids and their potential impactors) can be detected.  While the visibility of ejected dust in these cases is not expected to be necessarily correlated with orbit position (as in the case of sublimation-generated dust emission), consideration of other factors could help to maximize the likelihood of detecting dust features, such as timing observations to coincide with orbit plane crossings \citep[e.g.,][]{hsieh2005_phaethon,arendt2014_dirbedusttrails}, when the optical depth of a dust trail should be maximized, or close approaches to the Earth, when any dust should simply have a higher apparent surface brightness and thus be more easily detected.

The observational approach presented here could also be used to characterize dust features associated with large, bright asteroids suspected to have experienced recent impacts, such as (596) Scheila or (493) Griseldis (Section~\ref{subsection:background}).  While the initial detection of dust emission from these objects will usually presumably have been made using more conventional observing techniques, observations similar to those described here could help with the morphological characterization of close-in dust features that would otherwise be impossible to identify amid the large amount of scattered light from such bright objects, or extend the length of time over which fading dust features can be observed following an impact event.
Such observations could aid in setting limits on maximum ejected dust grain sizes, total ejected dust masses, and impactor sizes and energies, and also aid in the determination of the cause of the active event itself \citep[e.g.,][]{ishiguro2011_scheila2,ishiguro2011_scheila1}. 




\section{Summary and Conclusions}

In this paper, we report the results of an attempt to detect sublimation-driven dust emission from the large main-belt asteroid (24) Themis, which has been reported in previous studies to have widespread water ice frost on its surface, and whose collisional family contains at least 3 MBCs.  While no dust emission was ultimately detected, we suggest that our observational approach of using a chip gap to block much of the light from a bright target asteroid could be employed for future searches for faint dust features associated with Themis or other bright asteroids suspected of potentially exhibiting dust emission for one reason or another.

The observing strategy for the search described here involved searching for an extended dust trail aligned with the orbit plane of the object. Such a dust trail could be relatively long-lasting and, if present, would be expected to extend well past the residual scattered light expected from the $m_V$$\,\sim\,$11.5~mag object being blocked by the GMOS chip gap.  More detailed consideration of dust ejection physics, however, indicates that the maximum ejectable particles by sublimation at perihelion from an object as large and massive as Themis were unlikely to still be observable in the GMOS field of view at the time of our observations.

Despite the non-detection of a dust trail in this particular work, the results of our data reduction show that our observational and data reduction procedures would have enabled us to detect faint dust emission as close as 20~arcsec from the object (cf.\ Figure~\ref{figure:trail_profile}), even in a crowded star field.  As such, we suggest that similar future searches for dust emission from Themis or other large asteroids could target the immediate vicinity of the central asteroid, where dust particles small enough to exceed the object's escape velocity are expected to be most easily seen.  Given that these small particles should be relatively short-lived, in cases where sublimation-driven dust emission is expected, such searches should also be conducted much closer to perihelion than was done in this particular study, i.e., closer to the assumed time of peak dust emission.  In cases where dust emission driven by other mechanisms (such as impacts) is expected, conducting searches near perihelion is not necessarily advantageous because such activity is not expected to be correlated with orbit position.  In these cases, searches for dust should still be conducted when the visibility of dust features is expected to be highest, such as during orbit plane crossings, during close approaches to the Earth, or following detected disruption events.






\section*{Acknowledgements}

We thank Gemini support astronomers K.\ Chiboucas and L.\ Fuhrman, and queue observers J.\ Chavez and E.\ Lee for their assistance in obtaining our observations, and Gemini Observatory, which is operated by the Association of Universities for Research in Astronomy, Inc., under a cooperative agreement with the NSF on behalf of the Gemini partnership: the National Science Foundation (United States), the National Research Council (Canada), CONICYT (Chile), Ministerio de Ciencia, Tecnolog\'{i}a e Innovaci\'{o}n Productiva (Argentina), and Minist\'{e}rio da Ci\^{e}ncia, Tecnologia e Inova\c{c}\~{a}o (Brazil), for providing the opportunity to conduct Fast Turnaround program observations (Program GN-2015A-FT-19).  We also thank N.\ Artemieva and D.\ O'Brien for useful discussion.  H.H.H.\ acknowledges support from NASA Planetary Astronomy grant NNX14AJ38G, NASA Solar System Observations grant NNX16AD68G, and NASA Early Career Fellowship grant 80NSSC18K0193.  This work also benefited from support by the International Space Science Institute, Bern, Switzerland, through the hosting and provision of financial support for an international team, led by C.\ Snodgrass and including HHH and AF, to discuss the science of MBCs.

\acknowledgments

%

\vspace{5mm}
\facilities{Gemini-N(GMOS-N)}
\software{IRAF \citep{tody1986_iraf,tody1993_iraf}, Hotpants \citep{becker2015_hotpants}, WCSTools \citep{mink2011_wcstools}, SExtractor \citep{bertin1996_sextractor}}

\bibliographystyle{aasjournal}
\bibliography{hhsieh_refs}   



\end{document}